\documentclass[a4paper,11pt]{article}
\pdfoutput=1 
\usepackage{jcappub} 

\usepackage[T1]{fontenc} 
\usepackage{amsmath}    
\usepackage{graphicx}   
\usepackage{subcaption} 
\usepackage{verbatim}   
\usepackage{color}      
\usepackage{hyperref}   
\usepackage{footnote}
\usepackage{slashed}
\usepackage{amsfonts}
\usepackage{adjustbox}  

\newcommand{\be}{\begin{equation}}
\newcommand{\ee}{\end{equation}}
\newcommand{\bea}{\begin{eqnarray}}
\newcommand{\eea}{\end{eqnarray}}
\newcommand{\nn}{\nonumber}

\title{\boldmath Mirror Dirac leptogenesis}

\author[a]{K. Earl,}
\author[b]{C. S. Fong,}
\author[a]{T. Gregoire}
\author[a,1]{and A. Tonero\note{Corresponding author.}}

\affiliation[a]{Ottawa-Carleton, Institute for Physics, Carleton University,\\1125 Colonel By Drive, Ottawa, ON, K1S 5B6, Canada}
\affiliation[b]{Centro de Ci\^{e}ncias Naturais e Humanas, 
Universidade Federal do ABC,\\ Santo Andr\'{e}, 09210-580 SP, Brazil}

\emailAdd{kearl@physics.carleton.ca}
\emailAdd{cheesheng.fong@gmail.com}
\emailAdd{gregoire@physics.carleton.ca}
\emailAdd{alberto.tonero@gmail.com}

\abstract{ We consider a mirror world  scenario, in which light {\it Dirac neutrinos} are generated from a seesaw mechanism and leptogenesis occurs at high scale without violating lepton number. After leptogenesis, the conservation laws of the theory imply the visible baryon-minus-lepton asymmetry to be equal to the mirror baryon-minus-lepton asymmetry. We extend previous work by presenting a detailed study of this Dirac leptogenesis mechanism by constructing the full set of Boltzmann Equations (BEs) for both cases of unflavored and flavored regimes. We show that $Z_2$ breaking and lepton/mirror lepton flavor effects can be exploited to enhance the final baryon-minus-lepton asymmetry in our world by several orders of magnitude.
}

\begin{document}

\maketitle
\flushbottom

\section{Introduction}
The standard leptogenesis paradigm~\cite{Fukugita:1986hr} is characterized by lepton number violating decays of heavy Majorana singlets that occur out-of-equilibrium in the early universe and generate a primordial asymmetry in the lepton sector. Due to unsuppressed electroweak (EW) sphaleron reactions at high temperature, this lepton asymmetry is then transformed into a net baryon asymmetry. At the same time, the large Majorana mass scale elegantly gives rise to light Majorana Standard Model (SM) neutrinos through the type-I seesaw mechanism~\cite{Minkowski:1977sc,GellMann:1980vs,Yanagida:1979as,Mohapatra:1980yp}. 

The existence of a mirror world connected to the SM through some portal interactions has been widely discussed in the literature\footnote{See \cite{Blinnikov:1982eh,Blinnikov:1983gh,Khlopov:1989fj} for a few early studies.} and can be motivated in different ways. For example, it could be used to have a common origin for the baryon asymmetry and dark matter (DM). It could also help to solve the little hierarchy problem such as in Twin Higgs (TH) models~\cite{Chacko:2005pe}, in which the Higgs particle is a pseudo Nambu-Goldstone boson related to the spontaneous breaking of a global $SU(4)$ symmetry. The connection between baryon and mirror baryon asymmetries in these kinds of models as the consequence of a conserved baryon minus mirror baryon number has been discussed in~\cite{Farina:2015uea,Garcia:2015toa,Farina:2016ndq}. 

Another interesting class of models are those in which the SM is connected to the mirror sector through heavy Majorana singlets and light neutrino masses are generated in the two sectors through the type-I seesaw mechanism. Furthermore, decays of the heavy singlets produce lepton asymmetries in both sectors which then get converted into baryon and mirror baryon asymmetries~\cite{An:2009vq,Cui:2011wk}. Baryon-minus-lepton asymmetries of both sectors can be quite different depending on the parameters of the models.\footnote{Mirror symmetry can be invoked to enforce their equality~\cite{An:2009vq}. See also ref.~\cite{Falkowski:2011xh} where the authors considered heavy Majorana singlets which couple to a dark sector consisting of only a singlet scalar and fermion.} In these models, if the reheating temperature is lower than the mass of the heavy singlets, leptogenesis can also proceed through a different mechanism in which out-of-equilibrium, CP-violating scattering processes convert SM particles into particles of the mirror sector~\cite{Bento:2001rc}.

Although light neutrinos are required to explain the oscillation phenomena, their Majorana or Dirac nature can only be determined by dedicated lepton number violating experiments (like neutrinoless double beta decay) and this remains an open question. Therefore, one can ask if it is possible to envision a world in which the SM neutrinos are Dirac particles while the nice features of the seesaw mechanism and leptogenesis are preserved. 
Indeed, this possibility was first considered in~\cite{Gu:2012fg} with the introduction of heavy Dirac singlets and/or bidoublet Higgs charged under the SM and mirror EW symmetry.  
 The simplest scenario with the addition of heavy Dirac singlets is considered here. In this model
the SM neutrinos acquire tiny Dirac masses through Dirac type I seesaw where the role of right-handed neutrinos is played by the mirror neutrinos, and this implies that neutrinoless double beta decays are forbidden.
Dirac leptogenesis proceeds through decays of these heavy Dirac neutrinos to the SM leptons as well as the mirror leptons.\footnote{Earlier implementations of Dirac leptogenesis~\cite{Dick:1999je,Murayama:2002je} relied on the decays of heavy doublet particles to generate asymmetries in the left-handed lepton doublet and right-handed neutrino which are equal in magnitude and opposite in sign. The smallness of the neutrino masses generally implies that the left- and right-handed leptons are never in chemical equilibrium until much after EW sphaleron reactions are suppressed, in which part of the left-handed lepton asymmetry has already been converted to a net baryon asymmetry.} Interestingly, in this model the difference between baryon-minus-lepton numbers of the SM and the mirror sectors $(B-L)-(B'-L')$ remains a good symmetry of the theory (`prime' is used to indicate baryon and lepton numbers of the mirror sector). Starting from zero initial asymmetry, this conservation law implies that, after leptogenesis, the $B-L$ asymmetry has to be equal to the $B'-L'$ asymmetry, \emph{independently} of the details of the model.\footnote{This is similar in spirit to the hylogenesis mechanism where a global baryon number is imposed and dark matter is composed of dark anti-baryons \cite{Davoudiasl:2010am}.}
The final baryon and mirror baryon asymmetries, however, will be related by an order one coefficient, which depends on the details of the model.

If leptogenesis happens at high temperature\footnote{For the SM, such regime is $T \gtrsim 10^{12}$ GeV.} where lepton and mirror lepton flavors are not distinguishable (unflavored regime), CP violation is bounded from above~\cite{Gu:2012fg} analogous to the Davidson-Ibarra bound for type-I seesaw~\cite{Davidson:2002qv}. It was pointed out that this bound no longer holds once flavor effects are taken into account~\cite{Abada:2006fw,Nardi:2006fx,Abada:2006ea}. Therefore, extending the previous work~\cite{Gu:2012fg}, in this paper we will present a detailed study of the Dirac leptogenesis mechanism by constructing the full set of Boltzmann Equations (BEs) for both cases of unflavored and flavored regimes. We will show that $Z_2$ breaking and lepton/mirror lepton flavor effects can change the final asymmetry by several orders of magnitude. An attractive feature of the model is that due to the conservation of total $(B-L) - (B'-L')$ charge, one can exploit the mirror lepton flavor effects to enhance the production of $B-L$ asymmetry in our world.

This paper is organized as follows. In section \ref{sec:model} we introduce the model, discuss its global symmetries and present the leptogenesis CP violating parameters and related bounds. In section \ref{sec:BEs} we derive the unflavored BEs for the system. In section \ref{sec:benchmark} we solve the BEs in the unflavored case and show the results for two benchmark scenarios to illustrate the effect of $Z_2$ breaking. 
In section \ref{sec:flavor}, we solve the BEs in the flavored case and demonstrate the mechanism where enhancement can be achieved from purely mirror lepton flavor effects.
In section \ref{sec:discussion} we conclude with a final discussion. This work is supplemented by three appendices: in appendix \ref{app:proof} we give the proof of the identity used to obtain the bound on the CP parameter, in appendix \ref{app:Z2_sym_solutions} we present the approximate analytical solutions to the unflavored BEs in the limit where both the SM and the mirror sector share the same couplings and in appendix \ref{app:flavored} we present the flavored BEs appropriate for studying leptogenesis at a lower scale. 

\section{The model}
\label{sec:model}
We consider a model that, in addition to the SM sector, is characterized by the presence of a mirror sector with the same structure and field content as the SM. Such constructions have been proposed for a variety of reasons. For example, the mirror sector might contain dark matter, or could help alleviate the little hierarchy problem in the context of Twin Higgs models. It is quite natural in those models to impose a $Z_2$ symmetry that interchanges the SM and mirror sector.\footnote{In the rest of the text, we will use mirror sector to refer to the SM copy in generic mirror world models while reserving twin sector for the SM copy in the TH models.} The $Z_2$ symmetry justifies the particle content of the mirror sector and makes all the masses, Yukawa couplings and gauge couplings equal to the SM ones. Furthermore, in Twin Higgs models, a partial $Z_2$ that relates the top Yukawa coupling to the mirror top Yukawa coupling is important for the success of the mechanism~\cite{Chacko:2005pe,Chacko:2005vw,Craig:2015pha}. However, cosmological and phenomenological observations put tight constraints on these scenarios~\cite{Chacko:2005pe,Barbieri:2005ri,Barbieri:2016zxn,Chacko:2018vss} which forces some amount of $Z_2$ breaking. In particular, constraints on the number of relativistic degrees of freedom during nucleosynthesis and CMB formation are particularly difficult to avoid. In fact, the model we are considering requires at least two flavors of light right-handed neutrinos which, if they maintain thermal equilibrium with the Standard Model sector, would be enough to be in contradiction with the bound on $\Delta N_{\text{eff}}$. Introducing a large amount of  $Z_2$ breaking for the Yukawa of the first and second generations, as well as in the QCD scales of the SM and mirror sectors leading to different thermal histories and a colder mirror sector, could solve this problem \cite{Chacko:2005pe, Barbieri:2005ri,Barbieri:2016zxn}.  This scenario also suggests heavier mirror baryons which could explain why, if the mirror baryons are the dark matter, $\Omega_\text{DM} \sim 5 \Omega_m$ \cite{Barbieri:2005ri,Barbieri:2016zxn}. A $Z_2$ breaking in the Yukawa sector can be engineered, for example, by using Frogatt-Nielsen fields that have $Z_2$ breaking vevs \cite{Barbieri:2005ri}. Furthermore, in Twin Higgs models, some $Z_2$ breaking is required in the Higgs sector as well to avoid large deviations to Higgs measurements. This breaking could be explicit or spontaneous \cite{Beauchesne:2015lva}, but should keep the SM Higgs mass and the mirror world Higgs mass within the same order of magnitude if we want the dark baryons to be dark matter. In Twin Higgs models with a low cutoff, this proximity of mass scales is radiatively stable. However in models like the ones we are considering where the cutoff is high and the hierarchy problem is not addressed, without an exact $Z_2$ symmetry this proximity of mass scales appears accidental as there are large contributions to the SM and mirror Higgs masses both from the UV cutoff of the theory and from the heavy neutrinos. Ultimately, this issue could be resolved in models where the hierarchy problem is fully addressed, for example in a supersymmetric extension of our setup, where radiative corrections to the Higgs mass are proportional to the SUSY breaking mass scale which would make contributions from the neutrino sector subdominant.  In the extreme $Z_2$ breaking case, it is possible to remove all the light generations from the low energy spectrum, leading to the so-called Fraternal Twin Higgs~\cite{Craig:2015pha}.  It is also possible, instead of breaking the $Z_2$ in the Yukawa couplings, to break the symmetry in the way the two sectors are reheated~\cite{Berezhiani:1995am,Chacko:2016hvu}. In such a setup, one has to ensure that reheating happens after the two sectors lose thermal contact.
Even if the qualitative features of our model remain the same regardless of the presence of a $Z_2$ symmetry, the breaking of this symmetry can help with the enhancement of the asymmetry and will be considered in the discussion of some benchmark realizations of our model.

In addition to SM fields and their mirror copies, we add heavy singlet neutrinos to both sectors with a Dirac mass term which serves as a portal between the two sectors~\cite{Gu:2012fg}:
\bea\label{lagr1}
{\cal L}&=&i\bar N_{Ri} \slashed{\partial} N_{Ri}+i\bar N'_{Ri} \slashed{\partial} N'_{Ri}
-M_i \bar N_{Ri}  N'^c_{Ri}
\nn\\
&&
-y_{\alpha j}\bar{l}_{L\alpha}\tilde{\Phi}N_{Rj}-y'_{\alpha j}\bar{l'}_{L\alpha}\tilde{\Phi}'N'_{Rj}+{\rm h.c.}
\eea
The fields $l_{L\alpha}$ and $\Phi$ are SM lepton and Higgs doublets charged under the SM EW $SU(2)_L \times U(1)_Y$ and $\tilde \Phi=i\sigma_2\Phi^*$, while $l_{L\alpha}'$ and $\Phi'$ are mirror lepton and Higgs doublets that transform under the mirror EW group $SU(2)_L' \times U(1)_Y'$ and $\tilde \Phi'=i\sigma_2\Phi'^*$. The fields $N_{Ri}$ and $N'_{Ri}$ are heavy fermions which are singlets under both the SM and the mirror EW group. The number of generations in the mirror sector is not fixed, though a minimum of two generations of mirror fermions as well as $N_{Ri}$ and $N'_{Ri}$ are required for leptogenesis and to explain the two observed neutrino squared mass differences. In eq.~\eqref{lagr1}, the Dirac mass term is responsible for connecting the SM and the mirror sector.
In addition, we have chosen, without loss of generality, the basis where the Dirac mass matrix $M$, the charged lepton and mirror lepton Yukawa couplings (not shown above) are real and diagonal. Other portal interactions might exist, such as a Higgs portal or photon-mirror photon mixing, but are more model dependent.

The structure of eq.~\eqref{lagr1} can be obtained by imposing a global $U(1)$ symmetry that can be identified with the total lepton number $L_{\rm tot}=L-L'$ which is an extended lepton number defined in terms of both SM lepton number ($L$) and mirror lepton number ($L'$). We have:
\be 
L_{\rm tot}(l_{L \alpha}) = L_{\rm tot}(N_{Ri} ) = -L_{\rm tot}(l'_{L\alpha})=-L_{\rm tot}(N'_{Ri}), \qquad L_{\rm tot}(\Phi)=L_{\rm tot}(\Phi')=0.
\ee
In this case, the SM and twin right-handed neutrinos combine to form heavy Dirac states
\be 
N=N_R+ (N'_R)^c.
\ee
With this definition, we can rewrite the model Lagrangian as follows
\bea\label{lagrN}
{\cal L}&=&i\bar N_{i} \slashed{\partial} N_{i}
-M_i \bar N_{i}  N_{i}-y_{\alpha j}\bar{l}_{\alpha}\tilde{\Phi}P_RN_{j}-y'_{\alpha j}\bar{l'}_{\alpha}\tilde{\Phi}'P_RN^c_{j}+{\rm h.c.}
\eea

\subsection{Global symmetries of the model}
\label{subsec:symmetries}
The complete model has five $U(1)$ symmetries: $U(1)_{B}$, $U(1)_{B'}$, $U(1)_{L_{\rm tot}}$, $U(1)_{Y}$ and $U(1)_{Y^{'}}$. The last two are gauge symmetries which are anomaly free while the first three have $SU(2)_L$ and $SU(2)_{L'}$  mixed anomalies.\footnote{The anomalies are $SU(2)_L^2-U(1)_{B}$, $SU(2)_{L'}^2-U(1)_{B'}$, $SU(2)_L^2-U(1)_{L_{\rm tot}}$ and $SU(2)_{L'}^2-U(1)_{L_{\rm tot}}$.} With these symmetries we can form an anomaly free linear combination $U(1)_{B-B'-L_{\rm tot}}$. This symmetry could be gauged and broken spontaneously in various ways. If $U(1)_{B-B'-L_{\rm tot}}$ is broken at a scale $\mu \ll M_i$, Majorana masses for $N$ and $N'$ of the order $\mu$ can be naturally generated via operators of the form $\phi N N$ and $\phi' N' N'$ where $\phi$ and $\phi'$ are scalar fields which carry two units of positive and negative $B-B'-L_{\rm tot}$ charge, respectively. Since Dirac leptogenesis happens at high scale $T \sim M_i$, it will proceed as in our proposal, though with the interesting possibility that $N$ and $N'$ can be populated through additional interactions in the model. If $U(1)_{B-B'-L_{\rm tot}}$ is instead broken at a scale $\mu \sim M_i$, it is possible to avoid large Majorana masses for $N$ and $N'$ from being generated by choosing appropriately the scalar field content. For instance, by introducing only scalar fields $\phi$ and $\phi'$ which carry one unit of positive and negative ${B-B'-L_{\rm tot}}$ charge, respectively, we will generate Majorana masses of the order of $\mu^2/\Lambda$ through effective operators like $\phi^2 N N$ and $\phi'^2 N' N'$, where $\Lambda \gg M_i$ is the EFT expansion scale. This second scenario also brings in another interesting possibility of achieving resonant leptogenesis \cite{Pilaftsis:2003gt,Pilaftsis:2004xx,Pilaftsis:2005rv} if $\mu^2/\Lambda$ happens to be of the order of the decay width of $N$ and $N'$. We will comment more on this possibility towards the end of this work.

Let us define the normalized number density for particle species $x$ to be $Y_{x}=\frac{n_{x}}{s}$, where $s = \frac{2\pi^2}{45} g_\star T^3$ is the total entropic density of the Universe with $g_\star$ the number of relativistic degrees of freedom of the Universe ($g_\star = 213.5$ assuming an exact copy of the SM in the mirror sector). For particle $x$ with quantum number $x_q$ under $U(1)_q$, we use $Y_q$ to denote the normalized charge asymmetry  
\be 
Y_q = \sum_x x_q Y_{\Delta x},
\ee
where $Y_{\Delta x}=Y_{x}-Y_{\bar x}$. With these definitions we can rewrite the conservation of the total $B-B'-L_{\rm tot}$ as follows
\be \label{eq:B-L_cons}
 \sum_\alpha Y_{\Delta_\alpha}-\sum_\alpha Y_{\Delta'_\alpha}-\sum_i Y_{\Delta N_i}={\rm constant}
\ee
where $\Delta_\alpha \equiv B/3-L_\alpha$ and $\Delta'_\alpha \equiv B'/3-L'_\alpha$ with $L_\alpha$ and $L'_\alpha$ respectively referring to the lepton and mirror lepton flavor charges.
This relation will be verified explicitly in the BEs in the following sections. Assuming zero initial $B-B'-L_{\rm tot}$ asymmetry, and that leptogenesis completes before EW and mirror EW sphalerons freeze out, we have that, after all $N_i$ and $\bar N_i$ decay, the $B-L$ asymmetry in the visible sector is equal to the $B'-L'$ asymmetry in the mirror sector
\be 
Y_{\Delta }=Y_{\Delta '}, \label{eq:BL_eq}
\ee
where $Y_{\Delta }=\sum_\alpha Y_{\Delta_\alpha}$ and $Y_{\Delta '}=\sum_\alpha Y_{\Delta'_\alpha}$. Notice that this equivalence is independent of the presence of ${Z}_2$ symmetry breaking terms. This result is enforced by the global symmetries of the theory and represents a robust prediction of the model.

The relation between the baryon asymmetry $Y_{B}$ and $Y_{\Delta}$ and the relation between the mirror baryon asymmetry $Y_{B'}$ and $Y_{\Delta '}$ depend on the relativistic degrees of freedom that are present at the EW and twin EW sphaleron freeze out. In general, we have
\be 
Y_{B}=\kappa\, Y_{\Delta}, \qquad\qquad Y_{B'}=\kappa'\, Y_{\Delta '}.
\label{eq:B_BL}
\ee
If $Z_2$ is exact, one would have $\kappa = \kappa'$. But since it is typically expected that the $Z_2$ should be at least slightly broken, we can have $\kappa \neq \kappa'$, resulting in slightly different amounts of $B$ and $B'$ asymmetries taking into consideration eq.~\eqref{eq:BL_eq}. If mirror baryons are the DM with comparable mass to the SM baryons, this will provide an elegant explanation as to why the DM has similar energy density to the SM baryons.

\subsection{Heavy $N$ decay and CP violation}
In the model considered in this study we can have the following decay processes involving the heavy neutrinos $N_i\to l_\alpha \Phi$, $N_i\to\bar l'_\alpha \bar\Phi'$ and its antiparticle $\bar N_i\to \bar l_\alpha \bar\Phi$, $\bar N_i\to l'_\alpha \Phi'$. For generic complex Yukawa couplings $y$ and $y'$ in eq.~\eqref{lagrN} we can have CP violation in the decays of the heavy neutrinos and this will imply non-zero $\Delta\Gamma(N_i)_\alpha\equiv\Gamma(N_i\to l_\alpha \Phi)-\Gamma(\bar N_i\to \bar l_\alpha \bar\Phi)$ in the visible sector and non-zero $\Delta\Gamma'(\bar N_i)_\alpha\equiv\Gamma(\bar N_i\to l'_\alpha \Phi')-\Gamma(N_i\to\bar l'_\alpha \bar\Phi')$ in the mirror sector. Let us define $\Gamma(N_i)\equiv\sum_\alpha  [\Gamma(N_i\to l_\alpha \Phi)+\Gamma(N_i\to\bar l'_\alpha \bar\Phi')]$ and $\Gamma(\bar N_i)\equiv\sum_\alpha  [\Gamma(\bar N_i\to \bar l_\alpha \bar\Phi)+\Gamma(\bar N_i\to l'_\alpha \Phi')]$. 
CPT conservation implies that
\be 
\Gamma(N_i)=\Gamma(\bar N_i) \equiv \Gamma_{N_i},
\ee
where
\be 
\Gamma_{N_i}=\frac{M_i}{16\pi}[(y^\dagger y)_{ii}+(y'^\dagger y')_{ii}].
\ee
It follows that
\be \label{deltagammaequiv}
\sum_\alpha \Delta\Gamma(N_i)_\alpha =\sum_\alpha \Delta\Gamma'(\bar N_i)_\alpha.
\ee

\begin{figure}[ht!]
\centering
\includegraphics[scale=0.48]{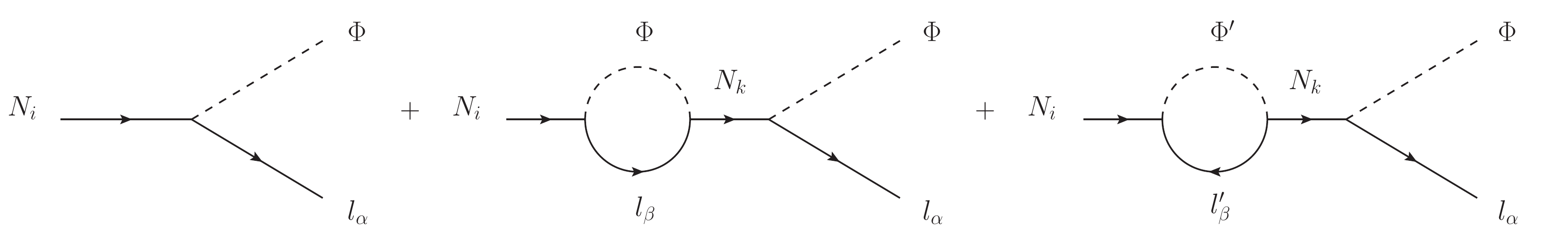} 
\includegraphics[scale=0.48]{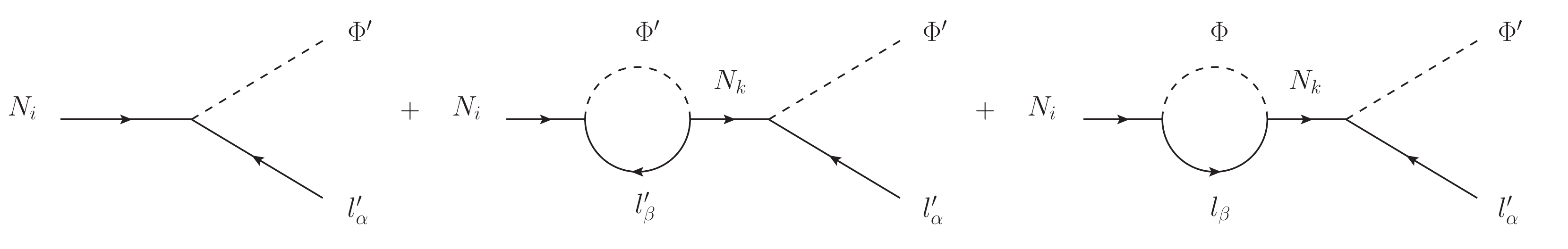} 
\caption{Diagrams responsible for the decay of the heavy neutrino $N$ into 
SM particles (top row) and the mirror particles (bottom row), contributing to the calculation of the CP violating parameters $\epsilon_{i\alpha}$ and $\epsilon_{i\alpha}^{'}$.} 
\label{NtoPL}
\end{figure}

Furthermore, we can define the CP violating parameters in 
the SM and mirror sector as follows
\bea 
\epsilon_{i\alpha} &=& \frac{\Delta\Gamma(N_i)_\alpha}{2\Gamma_{N_i}} , \\
\epsilon'_{i\alpha} &=& \frac{\Delta\Gamma'(\bar N_i)_\alpha}{2\Gamma_{N_i}}.
\eea
The relation in eq.~\eqref{deltagammaequiv} shows that the total CP violation in the visible and hidden sector is the same, namely
\be \label{eq:equalCP}
\epsilon_i = \epsilon_i' ,
\ee
where $\epsilon_i \equiv \sum_\alpha \epsilon_{i\alpha}$ and $\epsilon'_i \equiv \sum_\alpha \epsilon'_{i\alpha}$. 
The relation above is ensured by the presence of a global $B-B'-L_{\rm tot}$ symmetry as discussed in the previous section. The explicit computation of $\epsilon_{i\alpha}$ gives
\be 
\epsilon_{i\alpha}=\frac{1}{8\pi}\frac{1}{(y^{\dagger}y)_{ii}+(y^{'\dagger}y')_{ii}}\sum_{k}\left[\frac{1}{1-x_{k}}{\rm Im}[(y^{\dagger}y)_{ki}y_{\alpha k}y^*_{\alpha i}]+\frac{\sqrt{x_{k}}}{1-x_{k}}{\rm Im}[(y'^{\dagger}y')_{ik}y_{\alpha k}y^*_{\alpha i}]
\right],
\label{eq:ep_SM}
\ee
where $x_{k}=M_{k}^{2}/M_{i}^{2}$. This parameter is obtained from the interference between the tree- and loop-level diagrams in the top row of FIG.~\ref{NtoPL}. 
Interestingly, the right diagram on the top row of FIG.~\ref{NtoPL}
involves mirror particles in the loop and in performing the computation, ``the propagator of the internal $N_k$ picks up a mass term instead of the momentum" and is the only diagram that contributes in the unflavored case.

By a similar computation for the decay to the mirror sector (bottom row of FIG.~\ref{NtoPL}), 
we obtain $\epsilon'_{i\alpha}$
\be 
\epsilon'_{i\alpha}=\frac{1}{8\pi}\frac{1}{(y^{\dagger}y)_{ii}+(y^{'\dagger}y')_{ii}}\sum_{k}\left[\frac{1}{1-x_{k}}{\rm Im}[(y'^{\dagger}y')_{ki}y'_{\alpha k}y'^*_{\alpha i}]+\frac{\sqrt{x_{k}}}{1-x_{k}}{\rm Im}[(y^{\dagger}y)_{ik}y'_{\alpha k}y'^*_{\alpha i}]
\right].
\label{eq:ep_mirror}
\ee
Notice that in general the two CP parameter $\epsilon_{i\alpha}$ and $\epsilon'_{i\alpha}$ are different, however they  become equal in the $Z_2$ symmetric limit where $y=y'$. 

The middle diagrams of FIG.~\ref{NtoPL} involving the same type of particles in the loop as in the final states are relevant only in the flavored case. They give rise respectively to the first terms in the square brackets of eqs.~\eqref{eq:ep_SM} and \eqref{eq:ep_mirror} which vanish only if one sums over $\alpha$. As we will see in section \ref{sec:flavor}, this will provide a way to enhance the asymmetry generation utilizing the mirror lepton flavor effects.

Summing over $\alpha$, one can verify that eq.~\eqref{eq:equalCP} holds with
\be\label{epsilon}
\epsilon_i = \epsilon_i' =\frac{1}{8\pi}\frac{1}{(y^{\dagger}y)_{ii}+(y^{'\dagger}y')_{ii}}\sum_{k}\frac{\sqrt{x_{k}}}{1-x_{k}}{\rm Im}[(y'^{\dagger}y')_{ik}(y^{\dagger}y)_{ik}].
\ee
This parameter measures the total amount of CP violation induced by the decay of the heavy neutrino $N_i$. Compared to the standard leptogenesis result \cite{Fukugita:1986hr,Covi:1996wh}, there is no triangle or vertex diagram contribution and the CP violation comes only from the interference between the tree-level and the one-loop self-energy diagrams~\cite{Gu:2012fg}. 

\subsection{Neutrino masses and bound on CP violation}
In the limit of heavy right-handed neutrino masses, we can integrate out at tree level the $N$'s in eq.~\eqref{lagrN} by means of their equations of motion. Substituting back into the original Lagrangian we get the following dimension five effective operator involving the light neutrinos
\begin{equation}
{\cal L}_{\rm eff}=(yM^{-1}y'^{T})_{\alpha\beta}\bar{l}_{L\alpha}\tilde{\Phi}\tilde{\Phi}^{'T}(l_{L\beta}')^{c}+{\rm h.c.}
\end{equation}
After EW symmetry breaking in both sectors, we get the following Dirac mass term for the neutrinos
\begin{equation}
{\cal L}_{mass}=({\cal M}_\nu)_{ij}\,\bar{\nu}_{Li}\nu_{Rj}+{\rm h.c.}
\end{equation}
where the role of the right-handed neutrinos is taken by the mirror left-handed neutrinos, namely $\nu_{R}=(\nu'_{L})^{c}$. The explicit form of the mass matrix at leading order is given by the seesaw relation
\begin{equation}
{\cal M}_{\nu}=\upsilon f\,yM^{-1}y'^{T},
\label{eq:numass}
\end{equation}
where $\langle \tilde \Phi \rangle = (\upsilon\,\,0)^T$ with $\upsilon=174$ GeV and $\langle \tilde \Phi' \rangle = (f\,\,0)^T$, where $f$ is the VEV of the mirror Higgs doublet which is a free and model-dependent parameter. For instance, in the TH scenario, phenomenological constraints require $f \gtrsim 3 \upsilon$~\cite{Barbieri:2016zxn}. Disregarding the hierarchy problem and assuming $y \sim y' \sim 1$ and $f \sim 100\, \upsilon$, one can push the mass $M$ to the grand unification scale $\sim 10^{16}$ GeV in order to generate a neutrino mass of $0.1$ eV.

For definiteness, we will work with three generations of mirror leptons as well as $N_i$'s. In this case, we can parametrize the Yukawa matrices as follows
\bea 
y &=& \frac{1}{\sqrt{\upsilon f}}U^* D_{\sqrt{m}}X D_{\sqrt{M}},\\
y' &=& \frac{1}{\sqrt{\upsilon f}}V^*D_{\sqrt{m}}(X^{-1})^TD_{\sqrt{M}},
\eea
where $X$ is a $3 \times 3$ complex invertible square matrix, $D_{\sqrt{x}}$ is the square root of the diagonal matrix $D_x$ and $D_{M}\equiv M = {\rm diag}(M_1,M_2,M_3)$. The unitary matrices $U$ and $V$ are such that
\be 
U^T {\cal M}_{\nu} V ={\rm diag}(m_1,m_2,m_3) \equiv  D_m,
\ee
where $m_i$ are the physical light neutrino masses. Notice that this parametrization is a generalization of the parametrization introduced in~\cite{Casas:2001sr}.\footnote{The original Lagrangian contains a total of 18 + 3 moduli and 18 phases from $M$, $Y$ and $Y'$. The observables (in principle) are $2\times (3\, {\rm moduli} + 6\, {\rm phases})$ from $U$ and $V$, and 3 + 3 moduli from $D_m$ and $D_M$ giving a total of 12 moduli and 12 phases. The additional 9 moduli and 6 phases will be captured by the complex matrix $X$.}
For $M_1 \ll M_2,M_3$ we can write the CP parameter $\epsilon_1 $ in eq.~\eqref{epsilon} as follows
\be 
\epsilon_{1}
=-\frac{M_1}{8\pi}\frac{1}{\upsilon f}\frac{\sum_j m_j^2 {\rm Im}[(X^\dagger)_{1j} (X^{-1})^\dagger_{j1} ]}{\sum_j m_j (|X_{j1}|^2+|X^{-1}_{1j}|^2)}.
\ee
Using $X^{-1} X = \mathbf{1}_{3\times 3}$, we obtain the following inequality (see appendix \ref{app:proof} for details)\footnote{If the number of $N_i$ generations is equal to $k \neq 3$, in general, $X X^{-1}= \mathbf{1}_{3 \times 3}$ does not imply the condition $X^{-1} X = \mathbf{1}_{k \times k}$ required for the proof.}
\be \label{eq:twin_DI}
|\epsilon_1|\leq \frac{M_1 (m_3-m_1)}{16\pi}\frac{1}{\upsilon f}
= \frac{M_1 |\Delta m_{\rm atm}^2|}{16\pi (m_3 + m_1)}\frac{1}{\upsilon f} 
\equiv \epsilon_{1}^{\rm max},
\ee
where we have assumed $m_3 > m_2 > m_1$ and $|\Delta m_{\rm atm}^2|$ is the atmospheric square mass splitting. For the inverted mass ordering, we make the replacements $m_3 \to m_2$ and $m_1 \to m_3$. 
This relation is equivalent to the Davidson-Ibarra bound for the type-I seesaw~\cite{Davidson:2002qv} with the replacement $\frac{3}{\upsilon^{2}}\to\frac{1}{\upsilon f}$. If $f > \upsilon /3$, the lower bound on the mass of $M_1$ will be more stringent than the standard Davidson-Ibarra bound. Since $f$ is model-dependent, in principle, the bound can be relaxed by taking a small $f$.

\section{Boltzmann equations}
\label{sec:BEs}

In this section we will construct the BEs to describe the evolution of charge asymmetries $Y_{\Delta_\alpha}$, $Y_{\Delta_\alpha'}$ as well as heavy singlet densities $Y_{\Sigma N_i} \equiv Y_{N_i} + Y_{\bar N_i}$ and asymmetries $Y_{\Delta N_i}$.\footnote{For simplicity, we assume that both the EW and mirror EW sphaleron processes are in equilibrium and so the appropriate charge asymmetries to consider are $Y_{\Delta_\alpha}$ and $Y_{\Delta_\alpha'}$. For the SM, this is the case if leptogenesis occurs at $T \lesssim 2 \times 10^{12}$ GeV.} To focus on the important features, here we will present the BEs assuming leptogenesis proceeds through the decays of the lightest singlets $N_1$ and $\bar N_1$ and in the regime where both the SM lepton and mirror lepton flavors are not distinguishable.\footnote{Leptogenesis from decays of $N_2$ and $N_3$ can be neglected if we assume the reheating temperature is sufficiently below $M_2$ and $M_3$ or that the asymmetry generated is negligible due to strong washout and/or small CP parameters. For the SM, the lepton flavors are not distinguishable for $T \gtrsim 4\times 10^{11}$ GeV.} The complete BEs for the flavored case are presented in appendix \ref{app:flavored} and flavor effects will be studied in section \ref{sec:flavor}. Furthermore, we assume both the SM and mirror sector to have the same temperature and that mirror fermions also come in three generations.

The set of BEs are given by
\begin{eqnarray}
sHz\frac{dY_{\Sigma N_{1}}}{dz} & = & -\gamma_{N_{1}}\left(\frac{Y_{\Sigma N_{1}}}{Y_{N_{1}}^{{\rm {\rm eq}}}}-2\right) \label{eq:BE_sN}\\
sHz\frac{dY_{\Delta N_{1}}}{dz} & = & P_1\gamma_{N_{1}}\left(\frac{Y_{\Delta l}}{Y_{l}^{{\rm nor}}}+\frac{Y_{\Delta \Phi}}{Y_{\Phi}^{{\rm nor}}}-\frac{Y_{\Delta N_{1}}}{Y_{N_{1}}^{{\rm {\rm eq}}}}\right)
-P_1'\gamma_{N_{1}}\left(\frac{Y_{\Delta l^{'}}}{Y_{l^{'}}^{{\rm nor}}}+\frac{Y_{\Delta \Phi^{'}}}{Y_{\Phi^{'}}^{{\rm nor}}}+\frac{Y_{\Delta N_{1}}}{Y_{N_{1}}^{{\rm {\rm eq}}}}\right) \label{eq:BE_dN} \\
sHz\frac{dY_{\Delta}}{dz} & = & -\epsilon_{1}\gamma_{N_{1}}\left(\frac{Y_{\Sigma N_{1}}}{Y_{N_{1}}^{{\rm {\rm eq}}}}-2\right)
+P_1\gamma_{N_{1}}\left(\frac{Y_{\Delta l}}{Y_l^{{\rm nor}}}+\frac{Y_{\Delta \Phi}}{Y_{\Phi}^{{\rm nor}}}-\frac{Y_{\Delta N_{1}}}{Y_{N_{1}}^{{\rm {\rm eq}}}}\right) \label{eq:BE_Delta} \\
sHz\frac{dY_{\Delta'}}{dz} & = & -\epsilon_{1}^{'}\gamma_{N_{1}}\left(\frac{Y_{\Sigma N_{1}}}{Y_{N_{1}}^{{\rm {\rm eq}}}}-2\right)
+P_1'\gamma_{N_{1}}\left(\frac{Y_{\Delta l^{'}}}{Y_{l^{'}}^{{\rm nor}}}+\frac{Y_{\Delta \Phi^{'}}}{Y_{\Phi^{'}}^{{\rm nor}}}+\frac{Y_{\Delta N_{1}}}{Y_{N_{1}}^{{\rm {\rm eq}}}}\right)
\label{eq:BE_Deltap}
\end{eqnarray}
where $z\equiv \frac{M_1}{T}$, $H = 1.66\sqrt{g_\star}\frac{T^2}{M_{\rm Pl}}$ is the Hubble rate with $M_{\rm Pl} = 1.22 \times 10^{19}$ GeV, $Y_{N_1}^{\rm eq} = \frac{45}{4\pi^4 g_\star} z^2 {\cal K}_2(z)$ with ${\cal K}_m(z)$ the type-$m$ modified Bessel function of the second kind, and $Y_x^{\rm nor} \equiv \frac{15 g_x \zeta_x}{8\pi^2 g_\star}$ with $g_x$ the total degrees of freedom of $x$ and $\zeta_x = 1(2)$ for a  relativistic fermion (boson).\footnote{For the normalization of heavy particle $N_1$, we have used a Maxwell-Boltzmann distribution while the normalizations of other light (massless) particles take into account whether they are fermions or bosons (see appendix A of \cite{Fong:2011yx}).} Furthermore, $\gamma_{N_1}$ is the total decay reaction density of $N_1$ (which is equal to $\gamma_{\bar N_1}$ due to CPT) while $P_1$ and $P_1'$ are respectively the tree-level branching ratios for $N_1$ decays to $l \Phi$ and $\bar l' \bar\Phi'$ with $P_1 + P_1' = 1$.\footnote{For our calculation, we will use tree-level amplitudes and Maxwell-Boltzmann distributions which give $\gamma_{N_1} = \gamma_{\bar N_1} = s Y_{N_1}^{\rm eq} \Gamma_{N_1} \frac{{\cal K}_1(z)}{{\cal K}_2(z)} $.}

After identifying $U(1)$ charges and interactions in the thermal bath, we can write the particle asymmetries $Y_{\Delta l^{(')}} = A^{(')} Y_{\Delta^{(')}}$ and $Y_{\Delta H^{(')}} = C^{(')} Y_{\Delta^{(')}}$ \cite{Fong:2015vna}. 
Furthermore, from eqs.~\eqref{eq:BE_sN}--\eqref{eq:BE_Deltap}, one can verify explicitly that $U(1)_{B - B' - L_{\rm tot}}$ is conserved, i.e.~$\frac{d}{dz}\left(Y_{\Delta} - Y_{\Delta^{'}} - Y_{\Delta N_1}\right) = 0$, where we have made use of eq.~\eqref{eq:equalCP}. Assuming zero initial asymmetries, it follows that $Y_{\Delta N_1} = Y_{\Delta} - Y_{\Delta^{'}}$ (c.f.\ eq.~\eqref{eq:B-L_cons}). 
Using the relations above, eq.~\eqref{eq:equalCP} and $P_1' = 1 - P_1$, the BEs we need to solve are
\begin{eqnarray}\label{unflBE}
sHz\frac{dY_{\Sigma N_{1}}}{dz} & = & -\gamma_{N_{1}}\left(\frac{Y_{\Sigma N_{1}}}{Y_{N_{1}}^{{\rm {\rm eq}}}}-2\right)\nn\\
sHz\frac{dY_{\Delta}}{dz} & = & -\epsilon_{1}\gamma_{N_{1}}\left(\frac{Y_{\Sigma N_{1}}}{Y_{N_{1}}^{{\rm {\rm eq}}}}-2\right)
+P_1\gamma_{N_{1}}\left(
c \frac{Y_\Delta}{Y^{\rm nor}}
-\frac{Y_{\Delta} - Y_{\Delta^{'}}}{Y_{N_{1}}^{{\rm {\rm eq}}}}\right)\nn\\
sHz\frac{dY_{\Delta^{'}}}{dz} & = & -\epsilon_{1} \gamma_{N_{1}}\left(\frac{Y_{\Sigma N_{1}}}{Y_{N_{1}}^{{\rm {\rm eq}}}}-2\right)
+\left(1- P_1\right)\gamma_{N_{1}}\left(
c^{'} \frac{Y_\Delta^{'}}{Y^{\rm nor}}
+\frac{Y_{\Delta} - Y_{\Delta^{'}}}{Y_{N_{1}}^{{\rm {\rm eq}}}}\right)
\end{eqnarray}
where we have defined $Y^{\rm nor} \equiv \frac{15}{8\pi^2g_\star}$ and
\be
c^{(')} \equiv \frac{A^{(')}}{g_{l^{(')}}} + \frac{C^{(')}}{2 g_{\Phi^{(')}}}.
\ee
For the SM, $g_\ell = 2\times 3$ and $g_\Phi = 2$. The solutions of the BEs depend on the so-called washout parameter defined as
\be
K_1 \equiv \frac{\Gamma_{N_1}}{H(T=M_1)}
\equiv \frac{\tilde m_1}{m_0},
\ee
which characterizes the degree of which the decays of $N_1$ and $\bar N_1$ are out-of-equilibrium. In the second definition above, $\tilde m_1 \equiv \frac{(y^\dagger y)_{11} + (y'^\dagger y')_{11}}{M_1} \upsilon f$ 
and $m_0 \equiv 8.69\times10^{-3} \left(\frac{g_\star}{213.5}\right)^{1/2}
\left(\frac{f}{500\,{\rm GeV}}\right)\,{\rm eV}$. 
Notice that in principle $\tilde m_1$ which scales as $(y^\dagger y)_{11} + (y'^\dagger y')_{11}$ can be much larger than the neutrino mass which scales as $y y'$.

The final asymmetry can be parametrized as
\be
Y_{\Delta} = Y_{\Delta'} = -2 \epsilon_1\, \eta \, Y_{N_1}^{\rm eq}(z=0), 
\label{eq:YDelta_eta}
\ee
where $\eta$ is the so-called efficiency factor which is a function of $K_1$ and $Y_{N_1}^{\rm eq}(z=0) = \frac{45}{2\pi^4 g_\star}$. The most efficient case $\eta = 1$ is realized in the limit of weak washout $K_1 \ll 1$ and when one starts from thermal abundances of $N_1$ and $\bar N_1$. 

From eqs.~\eqref{eq:B_BL} and \eqref{eq:YDelta_eta} and using the maximal CP parameter eq.~\eqref{eq:twin_DI}, we can write down a lower bound for $M_1$ in terms of the observed baryon asymmetry $Y_{B}^{\rm obs}$ as
\bea
M_1 &\geq& \frac{16 \pi Y_{B}^{\rm obs}}
{2\kappa\eta Y_{N_1}^{\rm eq}(z=0)}
\frac{vf(m_3 + m_1)}{|\Delta m_{\rm atm}^2|} \nonumber \\
&=& 2.3\times 10^{10}\,{\rm GeV}
\left(\frac{f}{500\,{\rm GeV}}\right)
\left(\frac{1}{\eta}\right)
\left(\frac{30/97}{\kappa}\right)\nonumber \\
&& \times \left(\frac{Y_{B}^{\rm obs}}{8.7\times 10^{-10}}\right)
\left(\frac{m_3 + m_1}{0.1\,{\rm eV}}\right)
\left(\frac{2.5\times 10^{-3}{\rm eV}^2}{|\Delta m_{\rm atm}^2|}\right).
\label{eq:M1_bound}
\eea
The bound above applies strictly for the unflavored scenario with hierarchical $N_i$. Consideration of specific flavor alignment could relax the bound by a few orders of magnitudes \cite{Blanchet:2006be,Racker:2012vw} while having quasi-degenerate $N_i$ could further relax the bound down to sub-TeV scale through resonant enhancement \cite{Pilaftsis:2003gt,Pilaftsis:2004xx,Pilaftsis:2005rv}.

Finally, given that the observed ratio of DM and baryon energy densities is $r \approx 5.4$ and assuming all the DM to be the mirror baryons, from eq.~\eqref{eq:BL_eq}, the DM mass can be expressed as
\be
m_n' = 5.4 \left(\frac{r}{5.4}\right)
\left(\frac{\kappa}{\kappa'}\right) m_n,
\label{eq:DM_mass}
\ee
where $m_n \approx 1\,{\rm GeV}$ is the nucleon mass.

\section{$Z_2$ breaking effects}
\label{sec:benchmark}
In this section we will focus on the unflavored case presented in the previous section and discuss two different benchmark scenarios according to the presence or absence of a $Z_2$ symmetry in the neutrino sector. The $Z_2$ symmetry assumption, together with the size of the Yukawa couplings of the mirror sector influence the solutions of the BEs presented in the previous section.

\subsection{$Z_2$ symmetric case  $y=y'$} \label{sec:Z2_sym}
\begin{figure}[ht!]
\begin{center}
\includegraphics[scale=0.8]{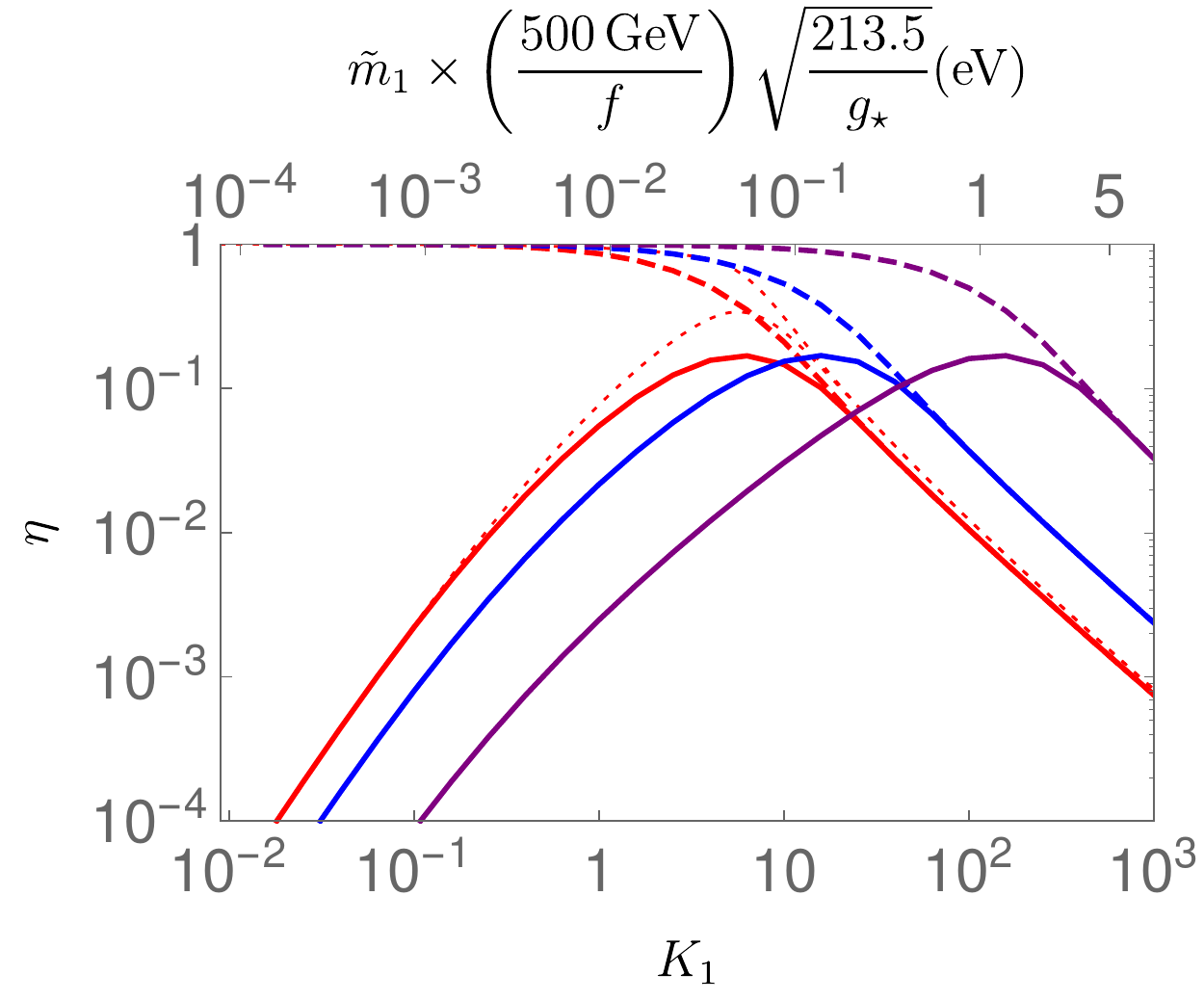} 
\caption{Efficiency factors as function of the washout parameter $K_1$ for $P_1 = 0.5$ (red), $P_1 = 0.9$ or $0.1$ (blue) and $P_1 = 0.99$ or $0.01$ (purple) for zero initial $N_1,\bar N_1$ abundances (solid lines) and thermal initial $N_1,\bar N_1$ abundances (dashed lines). The red dotted curves are the approximate solutions for $P_1 = 0.5$ presented in appendix \ref{app:Z2_sym_solutions} respectively for zero and thermal initial $N_1,\bar N_1$ abundances.
\label{fig:plot}}
\end{center}
\end{figure}

First we consider a scenario in which the SM and mirror neutrino sectors are related by an exact $Z_2$ symmetry, namely $y=y'$, and therefore $P_1 = \frac{1}{2}$. We consider the case for which $c = c^{'}$ which can be achieved either if the $Z_2$ symmetry is respected by the whole theory or if the $Z_2$ breaking in the quark and lepton sector is such that the differences in the Yukawa couplings do not affect the leptogenesis mechanism. Under these assumptions it can be shown that $Y_{\Delta} = Y_{\Delta '}$ at all times. For instance, assuming the first and second generation SM and mirror quark Yukawa interactions are not in equilibrium, we have $A = A^{'} = -\frac{3}{5}$ and $C = C^{'} = -\frac{2}{5}$. Therefore, we have
\be
c = c^{'} = -\frac{1}{5}. \label{eq:c_sym}
\ee
A good analytical approximate solution for the final asymptotic value of $\eta$ is presented in appendix \ref{app:Z2_sym_solutions}.
Finally, assuming (mirror) EW sphaleron processes freeze out after (mirror) EW symmetry breaking at a temperature below the (mirror) top mass, we have 
\be
\kappa = \kappa^{'} = \frac{30}{97}.\label{eq:kappa_sym}
\ee
In FIG.~\ref{fig:plot}, we show the efficiency factor $\eta$ as a function of the washout parameter $K_1$ for the $Z_2$ symmetric case with zero initial $N,\bar N$ abundances (red solid curve) and thermal initial $N,\bar N$ abundances (red dashed curve). The red dotted curves are the approximate solutions presented in appendix \ref{app:Z2_sym_solutions}.

\subsection{$Z_2$ broken case $y\neq y'$}
Next we consider a scenario in which the $Z_2$ symmetry is broken in the neutrino sector, namely $y\neq y'$, and therefore $P_1 \neq \frac{1}{2}$. Here we can have either that $c = c^{'}$ or $c \neq c^{'}$. During leptogenesis $Y_{\Delta} \neq Y_{\Delta '}$ and the equality is only established at the end of leptogenesis as in eq.~\eqref{eq:BL_eq}. 

As a concrete example, let us consider a scenario where we put the second generation mirror quark Yukawa interaction to be in equilibrium (due to larger mirror Yukawa coupling than the SM one), while other conditions remain the same as in the section \ref{sec:Z2_sym}. In this case the parameter $c$ has the same value as in eq.~\eqref{eq:c_sym}, while $c'$ with $A'=-\frac{3}{5}$ and $C'=-\frac{1}{4}$ is given by
\be
c' = -\frac{13}{80}.
\label{eq:c_diff}
\ee
In FIG.~\ref{fig:plot}, we show the efficiency $\eta$ as function of the washout parameter $K_1$ for $P_1 = 0.9$ or $0.1$ (blue curve) and $P_1 = 0.99$ or $0.01$ (purple curve) with zero initial $N,\bar N$ abundances (solid curves) and thermal initial $N,\bar N$ abundances (dashed curves). The choice of more extreme branching ratios induces a shift of the curve towards large values of $K_1$. The solutions assuming eq.~\eqref{eq:c_sym} or eq.~\eqref{eq:c_diff} essentially overlap due to the small difference between $c$ and $c'$. 

Finally, assuming mirror EW sphaleron processes freeze out after mirror EW symmetry breaking at a temperature below the mirror top and bottom masses, we have 
\be
\kappa^{'} = \frac{10}{41}.
\ee
In this case, the mass of dark matter will be equal to $5.4 \times \frac{30}{97} \times \frac{41}{10} = 6.8\, m_n$, which follows from eq.~\eqref{eq:DM_mass}.

\section{Flavor enhancement}
\label{sec:flavor}
Here we will discuss a novel enhancement effect that can be achieved in this model. Assuming that $|y'| \gg |y|$, from eqs.~\eqref{eq:ep_SM} and \eqref{eq:ep_mirror}, we have the following parametric dependence
\bea
\epsilon_{i \alpha} &\sim& a(y^4/y'^2) + b(y^2), \label{eq:epf}\\
\epsilon_{i \alpha}' &\sim& a(y'^2) + b(y^2), \label{eq:epfp}
\eea
where $a(x)$ represents the ``purely flavor terms'' i.e. the first terms in the square brackets of eqs.~\eqref{eq:ep_SM} and \eqref{eq:ep_mirror} while $b(x)$ represents the second terms in the square brackets of eqs.~\eqref{eq:ep_SM} and \eqref{eq:ep_mirror}.
Notice that the purely flavored term in the $\epsilon'$ parameter is enhanced by a factor of $\sim y'^2/y^2 \sim P^{-1}$ with $P$ the branching ratio for singlet decay to the SM sector.

For $N_1$ leptogenesis, the flavored terms will have an additional $M_1^2/M_{k>1}^2$ suppression while for $N_2$ leptogenesis, there can be an enhancement of $M_2^2/M_1^2$. Here will assume $M_1^2/M_{k>1}^2$ is of a factor of a few and focus on $N_1$ leptogenesis without resonant enhancement. If $N_1$ decays mostly to the mirror sector, the flavored CP parameters in the mirror sector can be enhanced accordingly. In this case, even if the flavored CP parameters in the SM are not enhanced, due to conservation of $(B-L)-(B'-L')$ charge, the enhanced production of mirror $(B'-L')$ asymmetry will be fed back to the SM sector, resulting in an overall enhancement of asymmetry production. This is an appealing feature of the model that we will explore in this section.

Since the flavored CP parameters in eq.~\eqref{eq:epfp} are enhanced by $ \sim r \equiv y'^2/y^2$, it becomes possible to reduce the lower bound on $M_1$ by a factor of $r$  compared to the unflavored bound eq.~\eqref{eq:M1_bound}. This is verified later in the section with a numerical example by solving the flavored BEs in eqs.~\eqref{eq:BEN}-\eqref{eq:BEdap}. Note that even with a very large $r$, $y'$ always remains in the perturbative regime due to the small neutrino mass in eq.~\eqref{eq:numass}. Despite $r \gg 1$, there is no tuning in order to obtain small neutrino mass which is proportional to $y y' = \sqrt{r} y^2$. One just has to choose corresponding smaller $y$. As compared to the type-I seesaw, choosing large neutrino Yukawa coupling while lowering the seesaw scale requires \emph{fine-tuned cancellations} in the neutrino mass matrix in order to maintain small neutrino mass.

If we keep increasing $r$, it seems like one can lower $M_1$ as much as we want. However, at some lower scale $M_1$, the flavor equilibrating scattering $l'_\alpha \Phi \leftrightarrow l'_\beta \Phi (\alpha \neq \beta)$ will be in equilibrium, making the flavor effect ineffective. To understand this, let us consider the following. If the flavor equilibrating scatterings are fast, one has $Y_{\Delta l'_e} = Y_{\Delta l'_\mu} = Y_{\Delta l'_\tau}$ which also implies $Y_{\Delta'_e} = Y_{\Delta'_\mu} = Y_{\Delta'_\tau}$.  Then, one would be able to sum up the BEs for $Y_{\Delta'_{\alpha}}$ as shown in eq.~\eqref{eq:BEdap}. As a result, the final $B'-L'$ asymmetry will be proportional to the total CP parameter $\epsilon_1'$. Since the result is independent on the flavored CP parameters eq.~\eqref{eq:epfp}, there is no flavor enhancement.

Next, let us estimate when flavor equilibrating scatterings become important. From eq.~\eqref{eq:ep_mirror}, the dominant flavor CP parameter in the mirror sector is given by
\bea
\epsilon_{1 \alpha}' &\sim& \frac{1}{8\pi}y'_{j\alpha} y'_{j\beta} \frac{M_1^2}{M_j^2}, \label{eq:epflap}
\eea
where we assume some hierarchy $(1-x_j)^{-1} \sim x_j^{-1}$ with $x_j \equiv M_j^2/M_1^2\, (j \neq 1)$ and $\beta$ is the flavor of coupling which is dominating. The scattering rate of $l'_\alpha \bar l'_\beta \leftrightarrow  \Phi \bar\Phi$ and $l'_\alpha \Phi \leftrightarrow l'_\beta \Phi \,(\alpha \neq \beta)$ through the off-shell exchange of $N_j (j=2,3)$ can be estimated by\footnote{The scattering through the exchange of $N_1$ is dominated by the on-shell contribution which is already taken into account by the decay and inverse decay of $N_1$.}
\bea
\Gamma_{\alpha\beta}' &\sim& \frac{1}{16\pi^3}(y'_{j\alpha} y'_{j\beta})^2 \frac{T^3}{M_j^2}. \label{eq:flaeq_scatt}
\eea
Requiring the scattering rate to be slower than the Hubble rate, we have
\bea
T \lesssim \frac{\pi}{4}\frac{1.66 \sqrt{g_\star}}{\epsilon_{1 \alpha}'^2} \frac{M_1^4}{M_j^2 M_{\rm Pl}}.
\eea
Taking $T \sim M_1$ and $g_\star = 213.5$, we have
\bea
M_1 \gtrsim 
6\times 10^6\, {\rm GeV} \left(\frac{\epsilon_{1 \alpha}'}{10^{-6}} \right)^2
\left(\frac{x_j}{10}\right).
\eea
The individual flavor CP parameters cannot be smaller than $10^{-6}$ to obtain successful leptogenesis and hence, $M_1$ cannot go much below $6\times 10^6$ GeV.\footnote{Similar generic bound for baryogenesis from decays was obtained in ref.~\cite{Sierra:2013kba}.} Below this scale, flavor equilibrating scatterings become important and at some point, flavor enhancement completely disappears.

Even if we are in the regime where lepton flavor equilibrating scatterings are not in equilibrium, flavor enhancement is ineffective if the branching ratio $P'_{1\alpha}$ is not hierarchical. In the limit of equal branching ratio  $P'_{1e} = P'_{1\mu} = P'_{1\tau} = 1/3$, one can again sum up the BEs for $Y_{\Delta'_{\alpha}}$ as shown in eq.~\eqref{eq:BEdap}. Therefore the final $B'-L'$ asymmetry will be proportional to the total CP parameter $\epsilon_1'$, rendering flavor enhancement ineffective. Finally, notice that the washout factor $K_1$ as well as branching ratio $P'_{1\alpha}$ depend on $y'_{1\alpha}$ while the mirror CP parameters eq.~\eqref{eq:epflap} and flavor equilibrating scatterings  eq.~\eqref{eq:flaeq_scatt} depend mainly on  $y'_{j\alpha}\,(j \neq 1)$. For instance, we can fix $K_1$ and $P'_{1\alpha}$ without affecting $\epsilon'_{1\alpha}$.

\begin{figure}[ht!]
\begin{center}
	\includegraphics[scale=0.6]{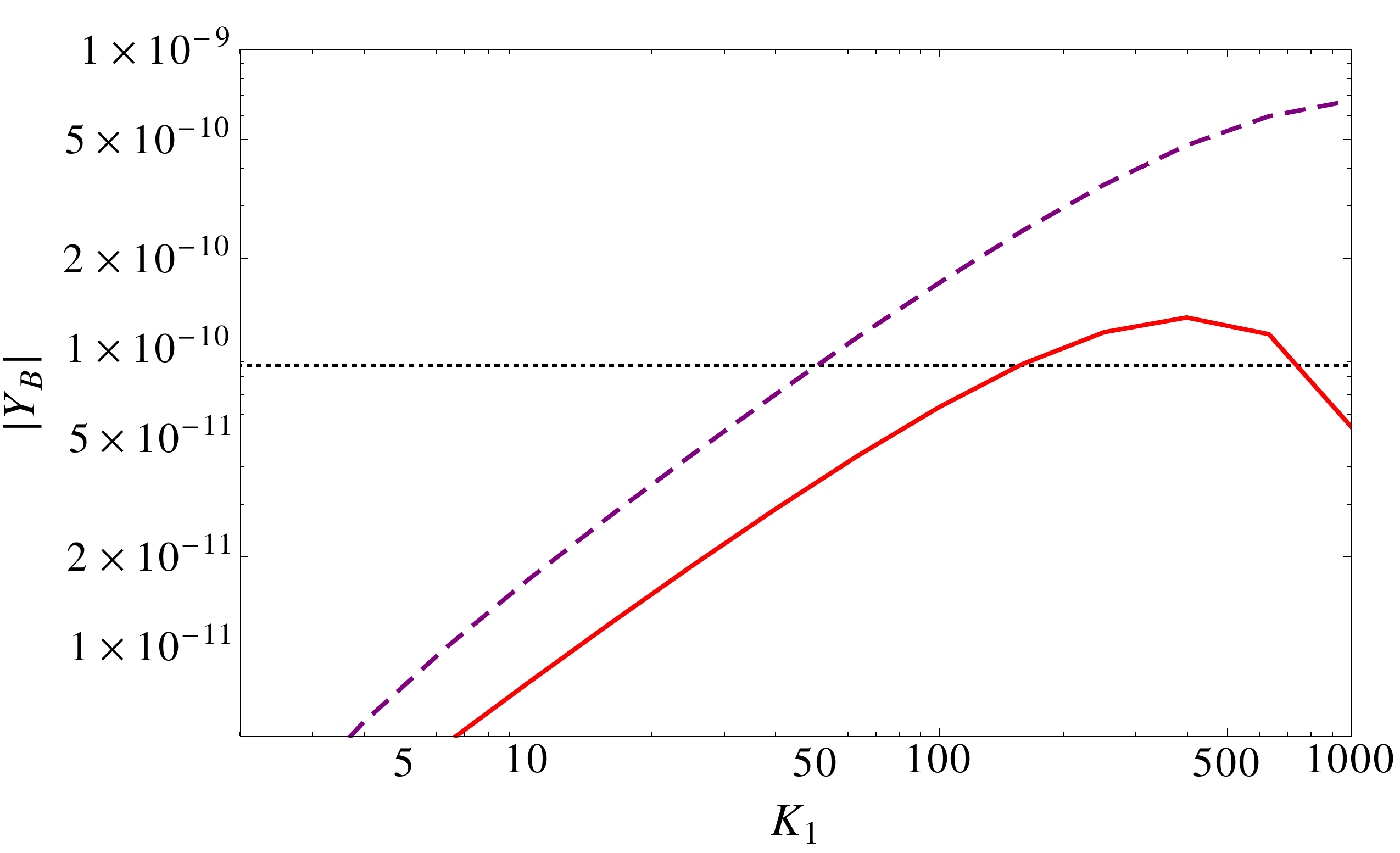} 
	\caption{The final baryon asymmetry $Y_B$ as a function of $K_1$ with $M_1 = 2\times 10^8$ GeV in the three-flavor regime assuming the parameters specified in eqs.~\eqref{eq:P1}-\eqref{eq:ep1p} and below them. The red solid and purple dashed lines represent respectively the baryon asymmetry obtained with zero and thermal initial $N_1$ abundance. The dotted horizontal line represents the value of the observed baryon asymmetry $Y_B = 8.7 \times 10^{-11}$. For the parameters specified above, the baryon asymmetry with initial thermal $N_1$ abundance has the wrong sign which could be changed by flipping the sign of CP parameters in eqs.~\eqref{eq:ep1} and \eqref{eq:ep1p}.
		\label{fig:plot_flavor}}
	\end{center}	
\end{figure}

As a concrete example, we illustrate the $N_1$ leptogenesis by choosing the following parameters with $r \simeq 10^3$
\bea
(P_{1e}, P_{1\mu}, P_{1\tau}) &=& 10^{-3}(1/3,1/3,1/3), \label{eq:P1} \\
(P_{1e}', P_{1\mu}', P_{1\tau}') &=& 0.999(8\times10^{-4},2\times10^{-4},0.999), \label{eq:P1p}\\
(\epsilon_{1e},\epsilon_{1\mu},\epsilon_{1\tau}) &=& -(1/3,1/3,1/3) \epsilon_1^{\rm max}, \label{eq:ep1} \\
(\epsilon_{1e}',\epsilon_{1\mu}',\epsilon_{1\tau}') &=& (1000,990,-1991) \epsilon_1^{\rm max} \label{eq:ep1p},
\eea
where the total CP parameter is $\epsilon_1 = \epsilon_1' = -\epsilon_1^{\rm max}$. By setting $M_1 = 2\times 10^8$ GeV, $f = 500$ GeV, $m_3 + m_1 = 0.1$ eV and the flavor matrices for both sectors to be eqs.~\eqref{eq:A1} and \eqref{eq:C1}, the final baryon asymmetry $Y_B$ obtained from solving the flavored BEs in eqs.~\eqref{eq:BEN}-\eqref{eq:BEdap} as a function of $K_1$ is plotted in FIG.~\ref{fig:plot_flavor}. The solid red and purple dashed lines refer respectively to the case with zero and thermal initial $N_1$ abundances while the dotted horizontal line is the observed baryon asymmetry $Y_B = 8.7\times 10^{-11}$. 

The democratic choice of parameters in eq.~\eqref{eq:P1} and \eqref{eq:ep1} are made such there is no flavor enhancement in our sector.
Regarding the flavor effects in the mirror sector, we have the following comments:
\begin{itemize}
\item The flavor effects are impotent in the weak washout regime because in this case the washout terms are negligible during decays of $N_1$, allowing one to sum over the source term and the final asymmetry will be proportional to the total CP parameter $\epsilon_1'$ which is too small in the three-flavor regime. For the parameters specified above, as shown in FIG.~\ref{fig:plot_flavor}, a sufficient baryon asymmetry can be generated for $K_1 \gtrsim 50$ and $K_1 \gtrsim 150$ respectively for thermal and zero initial $N_1$ abundance. 

\item For the zero initial $N_1$ abundance, the largest asymmetry is induced in the mirror flavor $\alpha$ for which $P'_{1\alpha} K_1 \gtrsim {\cal O}(1)$ in order to have a significant washout of the initial ``wrong'' sign baryon asymmetry generated during $N_1$ production. For our choice of parameters, the largest asymmetry is generated in the mirror $\tau$ sector.

\item On the other hand, for thermal initial $N_1$ abundance,  in the strong washout regime, the largest asymmetry is induced in the mirror flavor $\alpha$ for which $P'_{1\alpha} K_1$ is the smallest, i.e. the washout is the smallest.
For our choice of parameters, we have that the largest asymmetry is generated in the mirror $\mu$ sector. Moreover, for the parameters specified above, the final baryon asymmetry obtained with initial thermal $N_1$ abundance has the wrong sign compared to that with zero initial $N_1$ abundance. The correct sign can be obtained by flipping the signs of CP parameters in eqs.~\eqref{eq:ep1} and \eqref{eq:ep1p}.

\item  For $r \gg 1$, it is possible to reduce the scale of $M_1$ for successful leptogenesis by a factor of $r$ compared to the unflavored bound eq.~\eqref{eq:M1_bound}. We showed this in the example above where we took $r \sim 10^3$ and $P'_{1\alpha}$ to be hierarchical. To lower the scale further down to $10^{7}$ one can take $r\sim 10^4$. The flavor equilibrating scattering starts to become important at the scale below $6 \times 10^6$ GeV, diminishing the flavor enhancement. Therefore one cannot go much below that scale without resorting to resonant enhancement in the CP violation.

\item Since the model is symmetric under the exchange of $y \leftrightarrow y'$, we can also achieve the same enhancement by having $|y| \gg |y'|$.

\end{itemize}

\section{Discussion}
\label{sec:discussion}
In this work we have  considered a framework for leptogenesis from decays of heavy Dirac singlets 
which is characterized by the following features:
\begin{itemize}
\item assumes the existence of a mirror world with a global lepton number symmetry;
\item a seesaw mechanism generates small Dirac masses for the SM neutrinos which implies the absence of $0\nu \beta\beta$ decay;
\item leptogenesis occurs in a theory that respects a global lepton number symmetry and after leptogenesis has occurred, the symmetries of the theory enforce the $Y_{\Delta }$ asymmetry in the visible sector to be equal to the $Y_{\Delta'}$ asymmetry in the mirror sector;
\item the relation between $Y_{B}$ and $Y_{\Delta}$ as well as the relation between $Y_{B'}$ and $Y_{\Delta'}$ depend on the relativistic degrees of freedom that are present at the EW and mirror EW sphaleron freeze out, therefore the final baryon $Y_{B}$ and mirror baryon $Y_{B'}$ asymmetries are related by an order one coefficient, which depends on the details of the model;
\item for hierarchical $N_i$ and barring special flavor alignments, there exists a Davidson-Ibarra like bound on the CP parameter eq.~\eqref{eq:twin_DI} which in turn implies a lower bound on $M_1$ as given by eq.~\eqref{eq:M1_bound};
\item if mirror baryons are the DM, the model naturally describes an asymmetric dark matter scenario by providing an elegant explanation of why DM has similar energy density with the SM baryons.
\item The $Z_2$ breaking and flavor effects allow us to achieve enhanced production of asymmetry by a few orders of magnitude compared to the $Z_2$ symmetric and unflavored scenarios.
\end{itemize}

Though the qualitative features would be the same, it would be interesting to study in detail a scenario where the mirror sector starts with a colder temperature than the SM in view of the strong bounds on additional dark radiation. 

As we have seen, flavor effect can be utilized to lower the scale of leptogenesis by a few orders of magnitude down to $6\times 10^6$ GeV before flavor equilibrating scatterings render flavor enhancement ineffective.
For a variety of reasons, it might be desirable to further lower this scale down to sub-TeV. This can be achieved through resonant enhancement of CP violation by having quasi-degenerate $N_i$ which allows the circumvention of the bound given in eq.~\eqref{eq:twin_DI}. One possibility is to have resonant Dirac leptogenesis where one starts with quasi-degenerate Dirac mass for $N_i$. Another possibility, which could be realized quite naturally in models such as the one discussed in section~\ref{subsec:symmetries}, is to introduce small Majorana masses to split the Dirac fermions into quasi-degenerate Majorana fermion pairs. In either case, a mass splitting on the order of the decay width is required to have large enhancement. This kind of low scale leptogenesis can have a natural implementation in the framework of TH models.

\section*{Acknowledgments}
The authors would thank P. H. Gu for pointing out his earlier work on the subject.
The authors would also like to thank David Curtin and Andr\'e Lessa for useful discussions. This work was supported in part by the Natural Sciences and Engineering Research Council of Canada (NSERC). KE acknowledges support from the Ontario Graduate Scholarship (OGS).

\appendix
\section{Proof of the Davidson-Ibarra like bound}
\label{app:proof}
In this appendix, we provide a proof for the inequality 
\begin{align}
\Biggl| \frac{\sum_j m_j^2 \text{Im}[(X^\dagger)_{ij}(X^{-1})^\dagger_{ji}]}{\sum_j m_j (|X_{ji}|^2 + |X_{ij}^{-1}|^2)} \Biggr| \le \frac{1}{2}(m_3 - m_1),
\end{align}
which was used to derive eq.~\eqref{eq:twin_DI}. Here we assume the normal neutrino ordering so that $m_3 > m_2 > m_1$. First, it is convenient to introduce the notation $\text{Re}[X_{ij}^{-1}] = a_{ij}$, $\text{Im}[X_{ij}^{-1}] = b_{ij}$, $\text{Re}[X_{ji}] = c_{ji}$ and $\text{Im}[X_{ji}] = d_{ji}$. The above inequality can then be expressed in the following form
\begin{align}
\biggl| \sum_j m_j^2 (a_{ij}d_{ji} + b_{ij}c_{ji}) \biggr| \le \frac{1}{2}(m_3 - m_1) \sum_j m_j (a_{ij}^2 + b_{ij}^2 + c_{ji}^2 + d_{ji}^2). \label{eq:DI_US}
\end{align}
We will now show that eq.~\eqref{eq:DI_US} is valid. To do so, we first note that 
\begin{align}
1 = \sum_j X_{ij}^{-1} X_{ji} = \sum_j [a_{ij} c_{ji} - b_{ij}d_{ji} + i(a_{ij}d_{ji} + b_{ij}c_{ji})]
\end{align}
and therefore $\sum_j (a_{ij}d_{ji} + b_{ij}c_{ji}) = 0$. This, in turn, implies 
\begin{align}
\sum_{\substack{j \\ a_{ij}d_{ji} + b_{ij} c_{ji} > 0}} (a_{ij}d_{ji} + b_{ij}c_{ji}) = -\sum_{\substack{j \\ a_{ij}d_{ji} + b_{ij} c_{ji} < 0}} (a_{ij}d_{ji} + b_{ij}c_{ji}), \label{eq:DI_sumrule}
\end{align}
which we will use repeatedly. We introduce the notation
\begin{align}
\sum_{\substack{j \\ a_{ij}d_{ji} + b_{ij} c_{ji} > 0}} \equiv \sum_{j > 0} \qquad \text{and} \qquad \sum_{\substack{j \\ a_{ij}d_{ji} + b_{ij} c_{ji} < 0}} \equiv \sum_{j < 0}.
\end{align}
Next, we split the sum on the left-hand side of eq.~\eqref{eq:DI_US} into its positive and negative parts as follows
\begin{align}
\biggl| \sum_j m_j^2 (a_{ij}d_{ji} + b_{ij}c_{ji}) \biggr| = \biggl| \sum_{j > 0} m_j^2 (a_{ij}d_{ji} + b_{ij}c_{ji}) + \sum_{j < 0} m_j^2 (a_{ij}d_{ji} + b_{ij}c_{ji}) \biggr|. \label{eq:DI_step1}
\end{align}
Let us call the first sum on the right-hand side $S_+$ and the second sum on the right-hand side $S_-$, so that $S_+ > 0$ and $S_- < 0$. There are two cases we have to consider: $S_+ \ge |S_-|$ or $S_+ \le |S_-|$. If $S_+ \ge |S_-|$, it then follows that
\begin{align}
\biggl| \sum_j m_j^2 (a_{ij}d_{ji} + b_{ij}c_{ji}) \biggr| &= \sum_{j > 0} m_j^2 (a_{ij}d_{ji} + b_{ij}c_{ji}) + \sum_{j < 0} m_j^2 (a_{ij}d_{ji} + b_{ij}c_{ji}) \notag \\
&\le \sum_{j > 0} m_j^2 (a_{ij}d_{ji} + b_{ij}c_{ji}) + \sum_{j < 0} m_1^2 (a_{ij}d_{ji} + b_{ij}c_{ji}) \notag \\
&= \sum_{j > 0} m_j^2 (a_{ij}d_{ji} + b_{ij}c_{ji}) - \sum_{j > 0} m_1^2 (a_{ij}d_{ji} + b_{ij}c_{ji}) \notag \\
&\le (m_3 - m_1) \sum_{j > 0} (m_j + m_1) (a_{ij}d_{ji} + b_{ij}c_{ji}) \notag \\
&= (m_3 - m_1) \biggl( \sum_{j > 0} m_j (a_{ij}d_{ji} + b_{ij}c_{ji}) - \sum_{j < 0} m_1 (a_{ij}d_{ji} + b_{ij}c_{ji}) \biggr) \notag \\
&\le (m_3 - m_1) \biggl( \sum_{j > 0} m_j (a_{ij}d_{ji} + b_{ij}c_{ji}) + \sum_{j < 0} m_j |a_{ij}d_{ji} + b_{ij}c_{ji}| \biggr) \notag \\
&= (m_3 - m_1) \sum_{j} m_j |a_{ij}d_{ji} + b_{ij}c_{ji}|. \label{eq:DI_step2}
\end{align}
The proof for the case $S_+ \le |S_-|$ proceeds along the same lines and results in the same inequality as in \eqref{eq:DI_step2}.
Finally, we can use the triangle inequality and the fact that $|xy| \le (1/2)(x^2 + y^2)$ to conclude for both cases that 
\begin{align}
\biggl| \sum_j m_j^2 (a_{ij}d_{ji} + b_{ij}c_{ji}) \biggr| &\le (m_3 - m_1) \sum_{j} m_j (|a_{ij}d_{ji}| + |b_{ij}c_{ji}|) \notag \\
&\le \frac{1}{2}(m_3 - m_1) \sum_{j} m_j (a_{ij}^2 + b_{ij}^2 + c_{ji}^2 + d_{ji}^2). \label{eq:DI_step4}
\end{align}
This completes the proof. Notice that this proof can also be applied to the Davidson-Ibarra case \cite{Davidson:2002qv}. To do so, we simply need to identify the matrix $X$ with the (complex) orthogonal matrix $R$ of the Davidson-Ibarra case, and the matrix $X^{-1}$ with $R^T$. The property \eqref{eq:DI_sumrule} then becomes 
\begin{align*}
\sum_{\substack{j \\ \text{Im}(R_{ij}^2) > 0}} \text{Im}(R_{ij}^2) = - \sum_{\substack{j \\ \text{Im}(R_{ij}^2) < 0}} \text{Im}(R_{ij}^2)
\end{align*} 
and the proof follows in analogy to the more general case discussed above.

\section{Approximate solution for $\eta$ in the $Z_2$ symmetric case} \label{app:Z2_sym_solutions}
Here we write down the expression of the final efficiency factor $\eta$ obtained from an analytical approximate solution of the unflavored BEs of eq.~\eqref{unflBE}, in the limit of exact $Z_2$ symmetry where $P_1 = \frac{1}{2}$ and $c = c^{'}$ and $Y_{\Delta} = Y_{\Delta '}$ at all times. Assuming zero initial abundance of $N_1$ and $\bar N_1$, the final efficiency factor can be well approximated by\footnote{The detailed derivations can be found in refs.~\cite{Buchmuller:2004nz,Agashe:2018cuf}.}
\bea
\eta &=& -\frac{2}{\cal R}
e^{-\frac{3\pi}{8} {\cal R} K_1}
\left\{ 
\exp
\left[\frac{\frac{3\pi}{8} K_1}
{\left(1 + \sqrt{\frac{3\pi}{4}K_1}\right)^2}
{\cal R} \right] - 1
\right\} \nonumber \\ 
& & +
\frac{2}{z_B {\cal R} K_1}
\left\{ 1 - 
\exp
\left[-\frac{\frac{3\pi}{8} K_1}
{\left(1 + \sqrt{\frac{3\pi}{4}K_1}\right)^2}
z_B {\cal R} K_1
\right] 
\right\},
\eea
where ${\cal R} \equiv \frac{Y_{N_1}^{\rm eq}(z=0)}{Y^{\rm nor}}|c|$ and
\be
z_B = 1 + \frac{1}{2} 
\ln\left[1 + \frac{\pi K_1^2 {\cal R}^2}{1024}
\left(\ln\frac{3125\pi K_1^2 {\cal R}^2}{1024}\right)^5
\right].
\ee
For thermal initial abundance of $N_1$ and $\bar N_1$, the final efficiency is well approximated by
\be
\eta = \frac{2}{z_B {\cal R} K_1}
\left(1 - e^{-\frac{1}{2}z_B {\cal R} K_1}\right).
\ee

\section{Boltzmann equations in the flavored regime} \label{app:flavored}
Here we will provide the BEs where $\mu$ and $\tau$ flavored leptons and mirror leptons can be distinguished. These BEs are useful for studying leptogenesis at lower scales down to the EW symmetry breaking scale.\footnote{For the SM, this happens at $T\lesssim10^{9}$ GeV. Consideration of other possible flavor configurations for instance in the regime where only $\tau$ flavored lepton are distinguished while for the	mirror leptons, both $\mu$ and $\tau$ flavored can be distinguished is straightforward though less atheistic.} To be general, we include the contribution from all $N_{i}$, and we have
\begin{eqnarray}
	sHz\frac{dY_{\Sigma N_{i}}}{dz} & = & -\gamma_{N_{i}}\left(\frac{Y_{\Sigma N_{i}}}{Y_{N_{i}}^{{\rm {\rm eq}}}}-2\right),\label{eq:BEN}\\
	sHz\frac{dY_{\Delta N_{i}}}{dz} & = & \sum_{\alpha}\left[P_{i\alpha}\gamma_{N_{i}}\left(\sum_{\beta}c_{\alpha\beta}\frac{Y_{\Delta_{\beta}}}{Y^{{\rm nor}}}-\frac{Y_{\Delta N_{i}}}{Y_{N_{i}}^{{\rm {\rm eq}}}}\right)-P_{i\alpha}^{'}\gamma_{N_{i}}\left(\sum_{\beta}c_{\alpha\beta}^{'}\frac{Y_{\Delta_{\beta}^{'}}}{Y^{{\rm nor}}}+\frac{Y_{\Delta N_{i}}}{Y_{N_{i}}^{{\rm {\rm eq}}}}\right)\right],\label{eq:BEdN}\\
	sHz\frac{dY_{\Delta_{\alpha}}}{dz} & = & -\sum_{i}\left[\epsilon_{i\alpha}\gamma_{N_{i}}\left(\frac{Y_{\Sigma N_{1}}}{Y_{N_{1}}^{{\rm {\rm eq}}}}-2\right)-P_{i\alpha}\gamma_{N_{i}}\left(\sum_{\beta}c_{\alpha\beta}\frac{Y_{\Delta_{\beta}}}{Y^{{\rm nor}}}-\frac{Y_{\Delta N_{i}}}{Y_{N_{i}}^{{\rm {\rm eq}}}}\right)\right],\label{eq:BEda}\\
	sHz\frac{dY_{\Delta_{\alpha}^{'}}}{dz} & = & -\sum_{i}\left[\epsilon^{'}_{i\alpha}\gamma_{N_{i}}\left(\frac{Y_{\Sigma N_{1}}}{Y_{N_{1}}^{{\rm {\rm eq}}}}-2\right)-P_{i\alpha}^{'}\gamma_{N_{i}}\left(\sum_{\beta}c_{\alpha\beta}^{'}\frac{Y_{\Delta_{\beta}^{'}}}{Y^{{\rm nor}}}+\frac{Y_{\Delta N_{i}}}{Y_{N_{i}}^{{\rm {\rm eq}}}}\right)\right],\label{eq:BEdap}
\end{eqnarray}
where
\begin{eqnarray}
c_{\alpha\beta}^{(')} & \equiv & \frac{A_{\alpha\beta}^{(')}}{g_{l^{(')}}}+\frac{C_{\beta}^{(')}}{2g_{\Phi^{(')}}}\\
P_{i\alpha} & \equiv & \frac{\gamma\left(N_{i}\to l_{\alpha}\Phi\right)}{\gamma_{N_{i}}},\\
P_{i\alpha}^{'} & \equiv & \frac{\gamma\left(N_{i}\to\bar{l}_{\alpha}^{'}\bar{\Phi}^{'}\right)}{\gamma_{N_{i}}},
\end{eqnarray}
with $\sum_{\alpha}\left(P_{i\alpha}+P_{i\alpha}^{'}\right)=1$. For quantities $\gamma_{N_i}$ and $Y_{N_i}^{\rm eq}$, we have to make the replacement $z \to z M_i/M_1$. 

The values of $A_{\alpha\beta}$, $A_{\alpha\beta}^{'}$,
$C_{\alpha}$ and $C_{\alpha}^{'}$ depend on
the processes which are in chemical equilibrium. If leptogenesis takes
place in the temperature range $10^{7}\,{\rm GeV}\lesssim T\lesssim10^{9}\,{\rm GeV}$
where interactions mediated by up, down and electron Yukawa couplings
are out of equilibrium, we have
\begin{eqnarray}
A & = & \frac{1}{1074}\left(\begin{array}{ccc}
-906 & 120 & 120\\
75 & -688 & 28\\
75 & 28 & -688
\end{array}\right),\label{eq:A1} \\  
C & = & -\frac{1}{179}\left(37,52,52\right).  \label{eq:C1}
\end{eqnarray}
If leptogenesis takes place in the temperature range $10^{4}\,{\rm GeV}\lesssim T\lesssim10^{7}\,{\rm GeV}$
where only interactions mediated by the electron Yukawa coupling are out
of equilibrium, we have
\begin{eqnarray}
A & = & \frac{1}{1443}\left(\begin{array}{ccc}
-1221 & 156 & 156\\
111 & -910 & 52\\
111 & 52 & -910
\end{array}\right), \label{eq:A2} \\ 
C & = & -\frac{2}{481}\left(37,52,52\right).  \label{eq:C2}
\end{eqnarray}
If leptogenesis takes place in the temperature range $T\lesssim10^{4}$
GeV such that all processes mediated by Yukawa interactions are in
thermal equilibrium, we have
\begin{eqnarray}
A & = & \frac{2}{711}\left(\begin{array}{ccc}
-221 & 16 & 16\\
16 & -221 & 16\\
16 & 16 & -221
\end{array}\right), \label{eq:A3} \\ 
C & = & -\frac{16}{79}\left(1,1,1\right). \label{eq:C3}
\end{eqnarray}
If $Z_{2}$ is exact, i.e.\ the mirror sectors contain the same relativistic
degrees of freedom and the Yukawa couplings are exactly the same as
that of the SM, the matrices will be the same in the same temperature
range. Otherwise, they will not necessarily be the same. For instance, if mirror Yukawa couplings are larger, it is possible to have $A'$ and $C'$ as in eqs.~\eqref{eq:A3} and \eqref{eq:C3} while we are in the temperature regime $T \gtrsim 10^4$ GeV.

\end{document}